\documentclass[reprint,aps,prb,superscriptaddress,nofootinbib]{revtex4-2}
\usepackage[T1]{fontenc}
\usepackage{newtxtext,newtxmath}
\usepackage{graphicx}
\usepackage{siunitx}
\DeclareSIPrefix\micro{\ensuremath{\upmu}}{-6}
\usepackage{xspace}
\newcommand{\agentkerr}{\textsc{AgentKerr}\xspace}
\newcommand{\agentkerrlab}{\textsc{Agentic AI Kerr Lab}\xspace}
\usepackage{xurl}
\usepackage[colorlinks=true,linkcolor=blue!60!black,citecolor=blue!60!black,urlcolor=blue!60!black]{hyperref}
\hypersetup{
  pdftitle    = {The Agentic AI Kerr Lab: LLM Agents for Kerr Microscopy Experiments with Magnetic Skyrmions},
  pdfauthor   = {Kilian Leutner, Yuean Zhou, Aditya Kumar, Thomas B. Winkler, Robert Fr\"omter, Hans Fangohr, Mathias Kl\"aui},
  pdfkeywords = {magnetic skyrmions, Kerr microscopy, LLM agents, agentic AI, autonomous experiments, U-Net}
}
\newcommand{\figonewidth}{\textwidth}
\newenvironment{widefigure}{\begin{figure*}[t]}{\end{figure*}}
\newenvironment{acks}{\begin{acknowledgments}}{\end{acknowledgments}}
\newcommand{\figoneearly}{\figarchitecture}
\newcommand{\figonelate}{}
\begin{document}

\title{The Agentic AI Kerr Lab:\\ LLM Agents for Kerr Microscopy Experiments with Magnetic Skyrmions}

\author{Kilian~Leutner}
\email{kleutner@uni-mainz.de}
\affiliation{Institute of Physics, Johannes Gutenberg University Mainz, 55099 Mainz, Germany}
\author{Yuean~Zhou}
\affiliation{Institute of Physics, Johannes Gutenberg University Mainz, 55099 Mainz, Germany}
\affiliation{Max Planck Graduate Center Mainz, 55122 Mainz, Germany}
\author{Aditya~Kumar}
\email{kuaditya@uni-mainz.de}
\affiliation{Institute of Physics, Johannes Gutenberg University Mainz, 55099 Mainz, Germany}
\author{Thomas~B.~Winkler}
\affiliation{Institute for Molecules and Materials, Radboud University, 6525 AJ Nijmegen, The Netherlands}
\author{Robert~Fr\"omter}
\affiliation{Institute of Physics, Johannes Gutenberg University Mainz, 55099 Mainz, Germany}
\author{Hans~Fangohr}
\affiliation{Max Planck Institute for the Structure and Dynamics of Matter, 22761 Hamburg, Germany}
\affiliation{Center for Free-Electron Laser Science, 22607 Hamburg, Germany}
\affiliation{University of Southampton, Southampton SO17 1BJ, United Kingdom}
\author{Mathias~Kl\"aui}
\email{klaeui@uni-mainz.de}
\affiliation{Institute of Physics, Johannes Gutenberg University Mainz, 55099 Mainz, Germany}

\begin{abstract}
Magnetic skyrmions are topological spin textures that can be observed directly by magnetic microscopy and driven by fields and currents, which makes them a model system for fundamental physics and candidates for spintronic devices. Magneto-optical Kerr microscopy is a method to image them in real time; however, most experiments use an inefficient cycle of consecutive imaging, analysis and excitation in which every step requires the judgement of the experimentalist operating the setup. Here we present the \agentkerrlab (\agentkerr), in which an agent based on a large language model operates this cycle automatically and efficiently on a Kerr microscope. Rather than relying on pattern detection by a language model, a convolutional neural network trained on Kerr images turns live images into skyrmion observables such as position, size and shape, on which the agent bases its decisions. The architecture lets agents operate the instruments safely, as every hardware action is checked deterministically against an action policy, independently of the agent that issues it. With no procedure given, the agent determines by itself how to create, tune and move skyrmions, and writes its own analyses. The approach can be transferred to other spin textures and points towards magnetic imaging experiments in which the researcher poses the question and the agent finds the way to realize the experiments that yield the answer.

\end{abstract}

\maketitle
{\renewcommand{\thefootnote}{}\footnotetext{Contribution to the proceedings of the 2nd International Conference on AI for Science (AI4Sci 2026), held in Mainz, Germany, 5--9 October 2026.}}

\newcommand{\figarchitecture}{%
\begin{widefigure}
  \centering
  \includegraphics[width=\figonewidth]{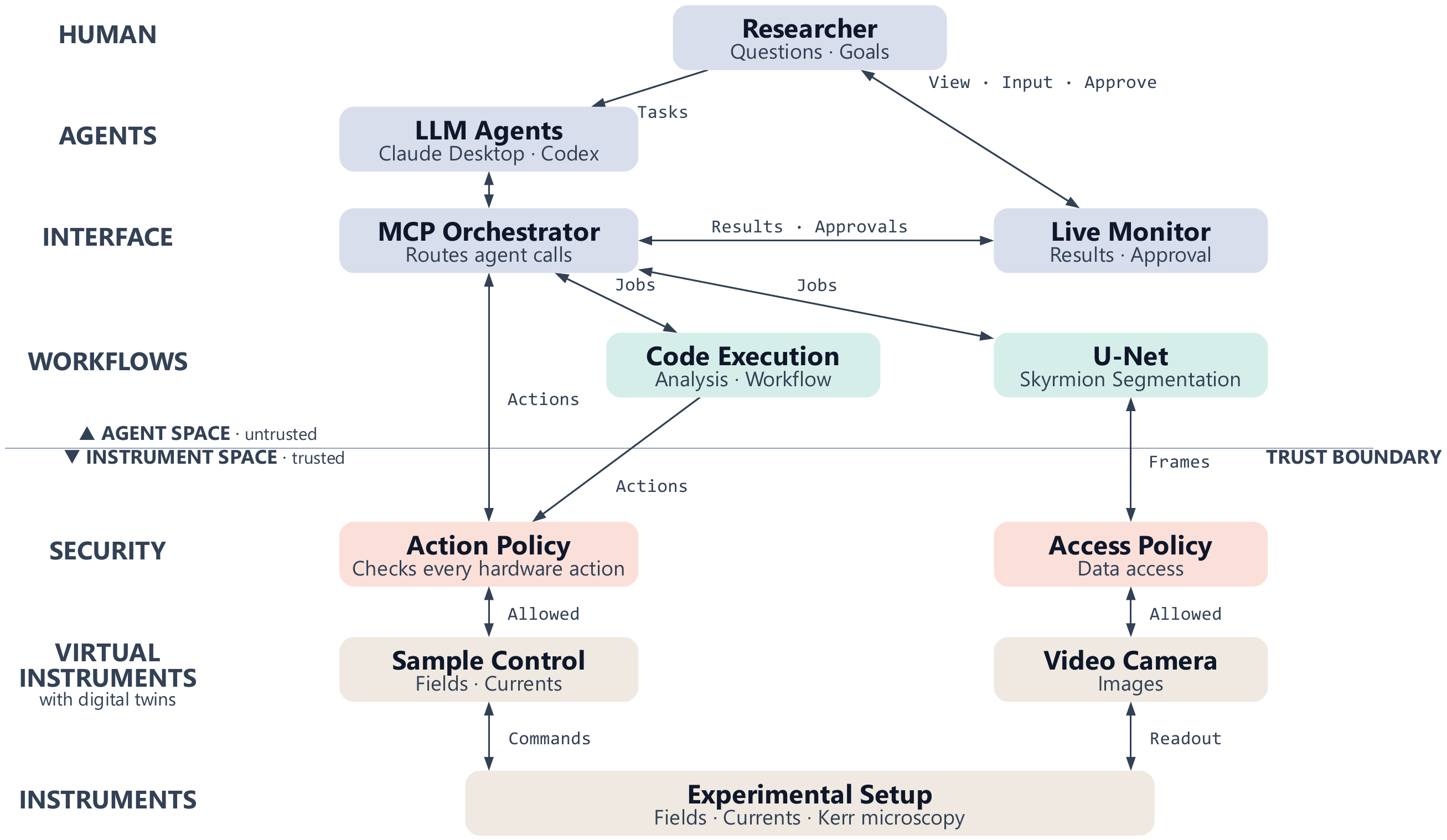}
  \caption{\textbf{Architecture of the \agentkerrlab.} The researcher states a goal, and LLM agents plan the experiment through the MCP orchestrator, with human interaction in the live monitor. On the workflow level, the U-Net segments the Kerr images and code execution runs analyses and complex experimental workflows. Below the trust boundary, the security layer checks all actions before virtual instruments apply them to the setup.}
  \label{fig:architecture}
\end{widefigure}}

Dipolar magnetic skyrmions are micron-sized, topologically stabilised, particle-like spin textures studied both for fundamental physics and for spintronic applications such as memory, sensing and unconventional computing~\cite{fert2017,bhukta2026,zazvorka2019,raab2022,beneke2024,leutner2023}. Magneto-optical Kerr microscopy, a table-top laboratory technique, images them directly and in real time~\cite{zazvorka2019,raab2022}. Experiments run in a cycle of imaging, analysis and excitation. Convolutional neural networks have automated the detection of skyrmions~\cite{labrieboulay2024}, but advanced experimental questions require elaborate analyses and a decision at every step, making the experimentalist operating the equipment often the bottleneck. Agents based on large language models (LLMs) combine reasoning with the use of external tools~\cite{yao2023react} and have begun to plan and execute laboratory work, from chemistry and self-driving laboratories~\cite{boiko2023,bran2024,muller2026optima} to instruments such as scanning probe, X-ray, electron, optical and quantum-sensing microscopes~\cite{mandal2025,slautin2026,chen2026,du2026,wall2026,yang2025,isogawa2026}, in several cases with image or measurement feedback. Connecting agents to specialised hardware, however, raises the question of how their actions are structured and controlled, and workflow and verification layers outside the language model have been proposed~\cite{isogawa2026,jamali2026,kang2026seguro,muller2026optima}.

This work introduces the \agentkerrlab (\agentkerr), in which an LLM agent closes this operating cycle on a magneto-optical Kerr microscope, as shown in Figs.~\ref{fig:architecture} and~\ref{fig:loop}: a convolutional neural network turns live images into quantitative skyrmion observables from which the agent chooses the next operation, a trust boundary keeps the safety of the hardware independent of the agent, and even instruments without a programming interface become accessible to agents and their workflows.

\figoneearly

\section{Quantitative perception of Kerr micrographs}

The experiments use a commercial Kerr microscope (evico magnetics) that records images at video rate, with out-of-plane and in-plane magnetic fields and electrical currents applied to the sample. Kerr images of skyrmions change from measurement to measurement in shape, motion blur, contrast, intensity gradients, defects and noise, so conventional image processing has to be redesigned by hand for each measurement and does not reliably separate skyrmions from defects. Instead, a convolutional U-Net~\cite{ronneberger2015} segments the live images pixel-wise into skyrmions, defects and background, where the background comprises magnetic and non-magnetic regions and their boundaries~\cite{labrieboulay2024,leutner2024thesis}; the model is trained on diverse skyrmions and defects on the micrometre length scale; where needed, the agent can adapt the pre- and post-processing. Running in PyTorch, with optimised post-processing that filters objects by size and shape, the segmentation reaches about 30 frames per second at $672 \times 512$ pixels on a GPU (NVIDIA GeForce RTX 2070) with about 0.5 GB of GPU memory, enabling real-time analysis. The agent therefore works with quantitative observables rather than raw pixels, which vision-language models still do not always analyse reliably~\cite{du2026}.

\section{An agentic magnetic microscope}

\figonelate

Building on the U-Net, \agentkerr operates the full cycle, as shown in Fig.~\ref{fig:architecture}: LLM agents, run in clients such as Codex or Claude Desktop, plan each experiment from the researcher's goal and reach \agentkerr through an orchestrator based on the Model Context Protocol (MCP)~\cite{mcp2024}, which routes all calls and provides a live monitor for user input. Since language-model outputs are inherently uncertain, the safety of the system does not depend on the agent: a trust boundary separates the agent space from the instrument space, where an action policy bounds experimental settings such as values, step sizes, waiting times and pulse durations by a fixed configuration, and, in the other direction, an access policy grants access to the measuring instruments only to specified entities. \agentkerr currently controls the microscope through the graphical user interface of its software: a researcher's mouse and keyboard input is recorded once, then modified and replayed as requested by the agent, verified by OCR and visual analysis; laboratories can thus connect agents to existing instruments without an API, while instruments with an API connect directly to \agentkerr.

Behind this security layer, virtual instruments receive the control actions and apply them to the real experiment; they are digital twins of the devices and include an emulator for testing. The video camera then records new images, and the U-Net returns the geometric shape and position of every skyrmion. These are passed to the agent directly or through the code execution tool, in which the agent writes its own analyses, such as ensemble statistics or tracking, including plots, before planning the next steps. Code execution and the U-Net form the workflow level above the trust boundary, so that all their hardware access also passes the security layer: the agent can plan complex workflows that are then executed deterministically in a sandbox with restricted external access.

\section{Agent-driven skyrmion experiments}

\begin{widefigure}
  \centering
  \includegraphics[width=\textwidth]{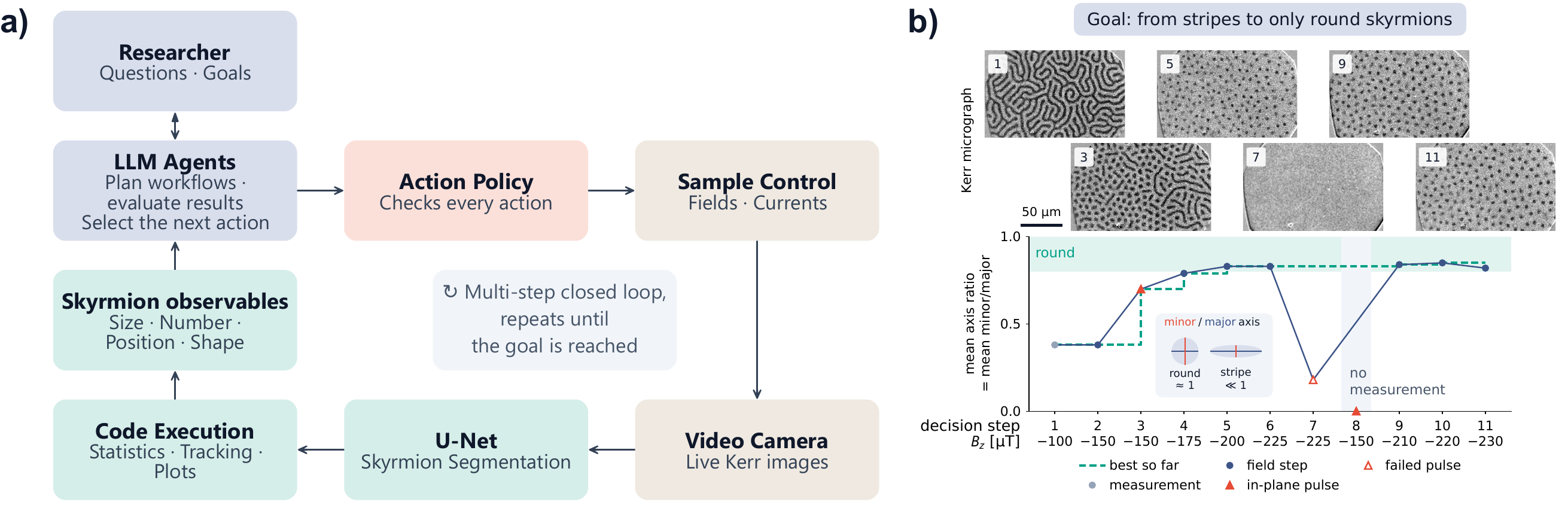}
  \caption{\textbf{The agentic loop and an example run.} a) The researcher asks a question; the agent runs the experiment and analyses it until the question is answered. b) As an example, one run from stripes to round skyrmions: measured mean axis ratio after each decision step, i.e. each pass through the loop. The agent chose the length of the in-plane pulses within safety limits; at step 7 a pulse annihilated the skyrmions, and the agent renucleated the state with skyrmions.}
  \label{fig:loop}
\end{widefigure}

With this architecture, the agent operates in the closed loop of Fig.~\ref{fig:loop}a. We ran agentic experiments with GPT-6 Astra, GPT-6 Luna and GPT-5.6 Luna as the underlying large language models; Fig.~\ref{fig:loop}b shows one representative run, in which the agent transforms a stripe-domain state into round skyrmions. Recordings of this run and of further experiments are available as a video on Zenodo~\cite{agentkerrvideos}: the nucleation of single skyrmions, the adjustment of the skyrmion radius to a target value, current-driven control, and analyses that the agent wrote on request, including advanced plots. No procedure was given for the run of Fig.~\ref{fig:loop}b, the radius adjustment or current-driven control; for single-skyrmion nucleation, the researcher outlined a strategy. As the state of a spin texture depends on the path of excitation, not only on the final values, the agent chose every step from the observed state and, for example at step 7 in Fig.~\ref{fig:loop}b, corrected its course after a failed step.

\section{Towards robust, unattended experiments}

At the moment, it is clear that this is not a finished project but \agentkerr is work in progress. Procedures the agent derived in these experiments are being captured as skills~\cite{agentskills2025}, human-readable instruction files loaded when a task requires them, so that the agent can carry out complex multi-step workflows such as field sweeps or nucleation protocols directly, without deriving them anew, and unattended. A further step is to connect the instrument through an API, which makes agent-based operation more direct than GUI automation. Future research directions include transferring the approach to other spin textures and magnetic imaging techniques and coupling it to micromagnetic simulations~\cite{abert2025} for insight into the underlying physics. Ultimately, experiments would require the researcher only to pose the question and to critically assess the result.

\begin{acks}
We acknowledge funding from ERC Synergy 3D-MAGiC (856538); Horizon Europe NIMFEIA (101070290) and MSCA TOPOCOM (101119608); DFG SPP 2137 Skyrmionics (403502522) and TRR 173 (268565370; A01, A12, B02); TopDyn, State of Rhineland-Palatinate; National Research Council of Science \& Technology (NST), Korean government MSIT (GTL24041-000).
\end{acks}

\bibliography{agentkerr}

\end{document}